\documentstyle[preprint,pra,aps]{revtex}

\begin{document}

\draft

\preprint{\vbox{
\null\hfill HUPD-9705 \\
\null\hfill AJC-HEP-30}}

\title{Measuring the two-photon decay width of Intermediate-mass Higgs
at a photon-photon collider}

\author{T.~Ohgaki\footnote{Corresponding author. E-mail address:
ohgaki@jlcux1.kek.jp} and T.~Takahashi}

\address{Department of Physics, Hiroshima University, \\
         1-3-1 Kagamiyama, Higashi-Hiroshima, 739, Japan}

\author{I.~Watanabe}

\address{Akita Junior College,\footnote{The name of the institute will 
         be changed to `Akita Keizai Hoka University Junior College'
         on April 1, 1997.} \\
         46-1 Morisawa, Sakura, Shimokitate, Akita, 010 Japan}

\date{\today}
\maketitle


\begin{abstract}

  Feasibility of a measurement of the partial decay width of the
intermediate-mass Higgs boson into two photons at a photon-photon
collider is studied by a simulation. The QCD radiative correction for 
quark pair background processes is taken into account for the realistic
background estimation. It is found that the two-photon decay width can
be measured with the statistical error of 7.6\% with about one year of 
experiment.  The impact of the measurement of the two-photon decay
width to look for the new physics beyond is demonstrated.

\end{abstract}


\narrowtext


\section{Introduction}
  
  The search and the study of Higgs boson, the last missing member of
the standard model family, are one of the most important tasks for the 
current and the future collider experiments at the energy frontier,
such as the CERN Large Electron-Positron Collider (LEP-II), the Next
Linear Collider (NLC), or the CERN Large Hadron Collider (LHC).

  The interaction of high energy photons at a photon-photon collider
\cite{gin81,gin83,tel90,tel95,wat93} provides us with an unique
opportunity to study the Higgs boson. The Higgs boson search at the
photon-photon collider has been studied by several authors
\cite{bor92,gun93,bor93,bor94,bai94,bai95,wat95,wat96,jik95,jik96,nlc96,jik93,naj89}.
Especially, it has been shown that the search for the
intermediate-mass Higgs in the mass range $M_{W} < M_{H} <2M_{W}$
through $\gamma\gamma \to H \to b\bar{b}$ process is complementary to
an $e^{+}e^{-}$ linear collider \cite{jon79} or a hadron collider
\cite{gun88,gun91,kun91}.

  Since two photons do not directly couple to the Higgs boson but
only do through loop diagrams of massive charged particles, any kind
of massive charged particles contribute to the two-photon decay width
of the Higgs boson if the mass of the loop particle is originated by
the Higgs mechanism \cite{gun90,oku82}. Figure~\ref{fig:cou} shows a
schematic diagram of the coupling of the Higgs boson with two
photons. It is notable that the contribution of a ultra-heavy particle
in the two-photon decay width of the Higgs boson does not been
suppressed but does keep a sizable constant if its mass is due to the
Higgs condensation.

  The deviation of the measured two-photon width from its predicted
value in the standard model (SM) indicates some additional
contributions from unknown particles, and thus it will be a signature
of new physics beyond SM which cannot be provided directly in the
ordinary collider experiments. For example, the supersymmetric
extensions of SM have additional charged particles such as scalar
fermions, charged Higgs and charginos. Since the masses of these new
particles partly originate from the Higgs mechanism of the electroweak
symmetry breaking, presence of these particles results in a shift of
the two-photon decay amplitude of the Higgs boson from its value of
SM. In fact, the minimal extension of the standard model (MSSM)
predicts the ratio of the two-photon decay widths $\Gamma(h^{0}
\to \gamma\gamma,\mbox{MSSM})/\Gamma(H \to \gamma\gamma, \mbox{SM})$
as much as 1.2 for the lightest Higgs boson with the mass of 120 GeV
\cite{bor93}.

  The intermediate-mass Higgs boson in SM mainly decays into a $b\bar{b}$
pair as is shown in Fig.~\ref{fig:bra}, and the daughter $b$-flavored
hadrons will be easily identified due to their long lifetime,
therefore the $b\bar{b}$ events are the best signals of the
intermediate-mass Higgs. The main background may be the continuum 
$\gamma\gamma \to q\bar{q}$ processes, however, the background events 
dominantly produced by initial photon collisions in $J_{z}=\pm 2$ angular 
momentum state can be suppressed by controlling the polarization 
of the colliding photon beams. Simultaneously, this control of the
beam polarizations causes to enhance the Higgs signals which are only 
accessible to the $J=0$ collisions \cite{gun93,bor93}. The feasibility
of the measurement of the two-photon decay width of Higgs boson in
this mass region have been studied using the Monte Carlo simulation by
Borden et al.\cite{bor92,bor93,bor94}. 

  Recently, several authors reported that the effect of QCD
corrections to $\gamma\gamma \to q\bar{q}$ is large since the helicity 
suppression which affects the background $q\bar{q}$ events does not
work due to a gluon emission. It could be a serious source of
backgrounds for the intermediate-mass Higgs, if some of the three-jet
events from $J_{z}=0$ state mimic two-jet events
\cite{bor94,jik95,jik96}. 

  In this work we simulate the measurement of the two-photon decay
width of the Higgs boson with the mass of 120 GeV at a future
photon-photon collider, including the effect of QCD corrections in the
manner of Jikia and Tkabladze \cite{jik96}. To perform a realistic
evaluation, the Monte Carlo programs CAIN \cite{che95,ohg95,che96},
JETSET 7.3 \cite{sjo94} and JLC-I detector simulator \cite{jlc92} are
applied for a luminosity distribution of a photon-photon collider,
hadronizations and selection performances in the detector,
respectively. The impact of the measurement for new physics search is 
discussed with estimated precision.


\section{Photon-Photon Collisions}

  We first summarize on the photon beam production, beam collision and 
luminosity distribution generated by CAIN simulation program
\cite{che95,ohg95,che96}. 

  As an example of a future linear collider, we adopt the parameters
of Japan Linear Collider (JLC) with X-band linear accelerators
\cite{jlc92}. In order to hit the mass pole of the Higgs boson at 120
GeV, the center-of-mass energy of the accelerator is tuned to be
$\sqrt{s}_{e^+e^-}$ =~150~GeV. We assume that the spent electrons are
bent away by the sweeping magnet so that only scattered photons
contribute to the luminosity. The energy of the laser photon is chosen
to be 4.18 eV, which results in the maximum photon energy to be
roughly 60 GeV. Parameters of the electron and laser beams are shown in
Table~\ref{tbl:par}. We assume the complete polarizations for both of
the electron and laser beams. The combination of the polarizations of
the electron $P_e$ and the laser $P_L$ should be $P_e P_L =-1.0$ so
that the generated photon spectrum peaks at its maximum energy. With
this combination of the electron and the laser beam polarizations, the 
obtained high energy photon beam is almost completely polarized around
the peak energy. 

  In order to enhance the Higgs production and to suppress the
background events, the polarizations of the colliding photon beams
should be arranged so that the $J_z=0$ collisions dominate. The
realistic luminosity distribution of the photon-photon collision is
provided by a Monte Carlo simulation program CAIN. CAIN is a
comprehensive simulation program of the Compton scatterings and of the 
beam-beam interactions between laser photons, electrons and positrons
in linear colliders. Figure~\ref{fig:lum1} shows the obtained
luminosity distribution of the photon-photon collider at
$\sqrt{s}_{e^+e^-}$=150 GeV. The $J_{z}=0$ and $\pm2$ components in
the luminosity distribution are plotted separately in
Figure~\ref{fig:lum1}. As mentioned above, $J_{z}=0$ component is
dominant in the luminosity distribution and occupies almost 100\%
around 120 GeV. Figure~\ref{fig:lum2} shows the luminosity
distribution in normalized c.m.s.~energy $z$ versus rapidity $\eta$
plane. Here, $z$ and $\eta$ are defined as, 
\begin{eqnarray}
\label{eq:1}
  z &=& \sqrt{s}_{\gamma\gamma}/2E_{e}=\sqrt{w_{1}w_{2}}/E_{e},\\
\label{eq:2}
  \eta &=& \log\sqrt{w_{1}/w_{2}},
\end{eqnarray} 
where $\sqrt{s}_{\gamma\gamma}$ is the $\gamma\gamma$ collision
energy, $E_{e}$ the energy of the electron beam, $w_{1}$ and $w_{2}$
the energies of left- and right-moving photons, respectively. It is
seen from the figure that the Higgs particle of 120 GeV is produced at
almost rest, and the low-energy background events like the resolved
photon processes are hardly boosted to have completely different
topologies from the signal events.


\section{Event Generation and Detector Simulation}

\subsection{Signal Events}

  For the intermediate-mass Higgs, the cross section of the process 
$\gamma\gamma \to H \to b\bar{b}$ near the mass pole can be described 
by the Breit--Wigner approximation,
\begin{eqnarray}
\label{eq:3}
  \sigma_{\gamma\gamma \to H \to b\bar{b}}=8\pi\frac{ \Gamma(H \to 
\gamma\gamma) \Gamma(H \to b\bar{b}) }{ (s_{\gamma\gamma}-M^{2}_{H})
^{2} + M^{2}_{H}\Gamma^{2}_{H} }(1+\lambda_1\lambda_2),
\end{eqnarray}
where $M_H$ is the Higgs mass, $\Gamma(H \to \gamma\gamma)$ and 
$\Gamma(H \to b\bar{b})$ the decay widths of the Higgs boson into two 
photons and $b$ quark pair, $\Gamma_{H}$ the total decay width,
$\lambda_1$ and $\lambda_2$ the initial photon helicities, 
respectively. 

  The total number of produced Higgs bosons is estimated by
convoluting the differential luminosity distribution calculated by
CAIN with Eq.(\ref{eq:3}). The effective cross section
$\sigma^{eff}_{|\cos\theta|<0.95}$ obtained by the convolution of
differential luminosity distribution with Eq.(\ref{eq:3}) is given in 
Table~\ref{tbl:crs}. A kinematical cut $|\cos\theta|<0.95$ for the
scattered angle $\theta$ of $b$ and $\bar{b}$ quarks in the
center-of-mass system of the colliding photons is imposed. Throughout
our analyses we adopt the quark masses of $m_{b}$=4.3 GeV, $m_{c}$=1.3
GeV, and $m_{t}$=176 GeV. The branching ratios $Br(H \to b\bar{b})$
and $Br(H \to \gamma\gamma)$ in SM are 64.3\% and 0.243\%,
respectively, which are computed by HDECAY program \cite{spi96}. The
number of events of the $b\bar{b}$ pairs from Higgs decay will be
5,080 for an integrated luminosity of 10 $\mbox{fb}^{-1}$ which
roughly corresponds to a one-year run. 

  For the further analyses of detector acceptance, four-momenta of $b$
and $\bar{b}$ from Higgs decay are generated by
BASES/SPRING\cite{kaw86}. Subsequent hadronizations of quarks are
simulated by the parton shower picture with JETSET 7.3 \cite{sjo94}. 

\subsection{Background Events}

  The $\gamma\gamma \to q\bar{q}$ background events are generated in
a similar way as in the Higgs production, except that the production
amplitudes are calculated by HELAS\cite{mur92}, and except that only 
the events with $\sqrt{s}_{\gamma\gamma}$ $>$~75~GeV are generated. 
The shapes of three-jet events are reproduced by a parton shower 
treatment of $q\bar{q}$ evolution by JETSET 7.3, and the QCD
corrections of the soft gluon emission, hard gluon emission, and
virtual correction to the cross section normalization are taken into
account {\it \`a la} Jikia and Tkabladze \cite{jik96}. 

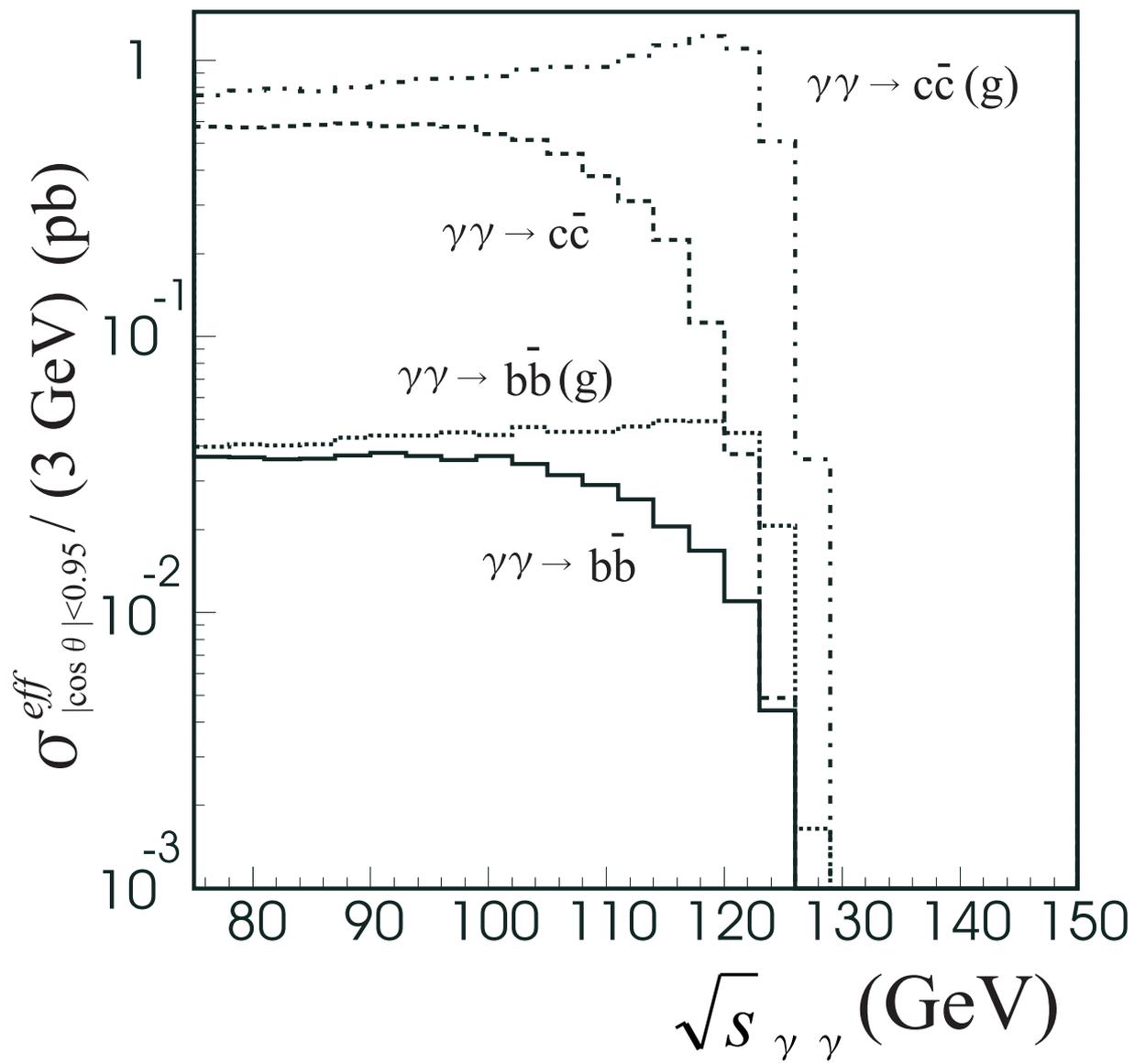
\begin{figure}
\caption{The collision energy distributions of the effective cross
sections at a photon-photon collider. The solid line corresponds to
the tree-level $\gamma\gamma \to b\bar{b}$, dashed line to tree-level 
$\gamma\gamma \to c\bar{c}$, dotted line to $\gamma\gamma \to
b\bar{b}$ with QCD corrections and 
dash-dotted line to $\gamma\gamma \to c\bar{c}$ with QCD
corrections. The bin size is 3~GeV.}
\label{fig:parton}
\end{figure}

  The effective cross sections and the number of the generated events 
of the background processes with and without the QCD corrections are 
also listed in Table~\ref{tbl:crs}. In this table, $\gamma\gamma \to
q\bar{q}(g)$ indicates the process $\gamma\gamma \to q\bar{q}$ taking 
account of the QCD corrections. Figure~\ref{fig:parton} shows the 
effective cross sections. From this figure, one finds that the QCD
correction is drastically large at the maximum collision energy, where
the tree $q\bar{q}$ production in $J_z=0$ mode is hardly suppressed by
the helicity conservation law. The effective cross section of
$\gamma\gamma \to c\bar{c}$ is larger than that of $\gamma\gamma \to
b\bar{b}$ due to the large electric charge of the quark. 

  In Table~\ref{tbl:crs}, we also listed the processes of
$\gamma\gamma \to H \to c\bar{c}$ and $\gamma\gamma \to H \to gg$ as 
backgrounds. The branching ratios of $Br(H \to c\bar{c})$ and $Br(H
\to gg)$ are set to be 2.67\% and 8.03\%, respectively \cite{spi96}.  

\subsection{Detector Simulation}

  In order to demonstrate the identification of the Higgs events at a 
photon-photon collider, we used the JLC detector simulation program
which smears the kinematics of the final-state particles according to 
the JLC-I detector resolution\cite{jlc92}. The performance parameters
of the JLC-I detector can be found in Table~\ref{tbl:jlc}. The main
components used in this simulator are the vertex detector, central
drift chamber and calorimeters. The $b$-quark tagging by the vertex
detector is crucial in this analysis. A CCD detector is assumed in the 
current JLC-I design, and its resolution of the impact parameter is,
\begin{eqnarray}
\label{eq:4}
  \sigma_{d}^2=11.4^{2}+(28.8/p)^{2}/\sin^{3}{\theta} \ 
(\mu\mbox{m}^2),
\end{eqnarray}
where $p$ is the momentum of the charged particle in GeV, $\theta$ is
the scattering angle.


\section{Event Analysis}

  The analysis requires the reconstruction of the two-jet final states
from Higgs boson decay. To identify the $b\bar{b}$ final states from
Higgs, we introduce some $b$-tagging requirements.  

  First of all, well reconstructed tracks and clusters in calorimeters 
are selected from the generated events by the Monte Carlo, and only
these tracks and clusters are used in the further analysis. A `good
track' is required $|\cos\theta|<0.95$, $P_{t}>0.1$ GeV and CDC-VTX
track matching. A `good cluster' is defined as $E>0.1$ GeV and
$|\cos\theta|<0.99$. 

  The number of good tracks is required to be greater than 10 to
choose multi-hadron events, and then two-jet events are selected by
JADE clustering algorithm\cite{bar84} with $y_{cut}$=0.02. A cut of 
$|\cos\theta_{jet}|<0.7$, where $\theta_{jet}$ is the scattering angle
of the jet, is applied to make sure that the events are well contained
in the detector volume and to increase the ratio of signal events to 
backgrounds. 

  A $b$ ($\bar{b}$) jet is selected by requiring that five or more
tracks which have the normalized impact parameter $d/\sigma_{d}>2.5$
and $d<1.0 \ \mbox{mm}$ are in each jet, where $d$ is the impact
parameter. Only the events that both of the two jets are tagged as
$b$-jets are regarded to be the $b\bar{b}$ events to improve the
reject probability of the charmed events. The resulting number of the
tagged events are summarized in Table~\ref{tbl:tag}. 


\section{Results}

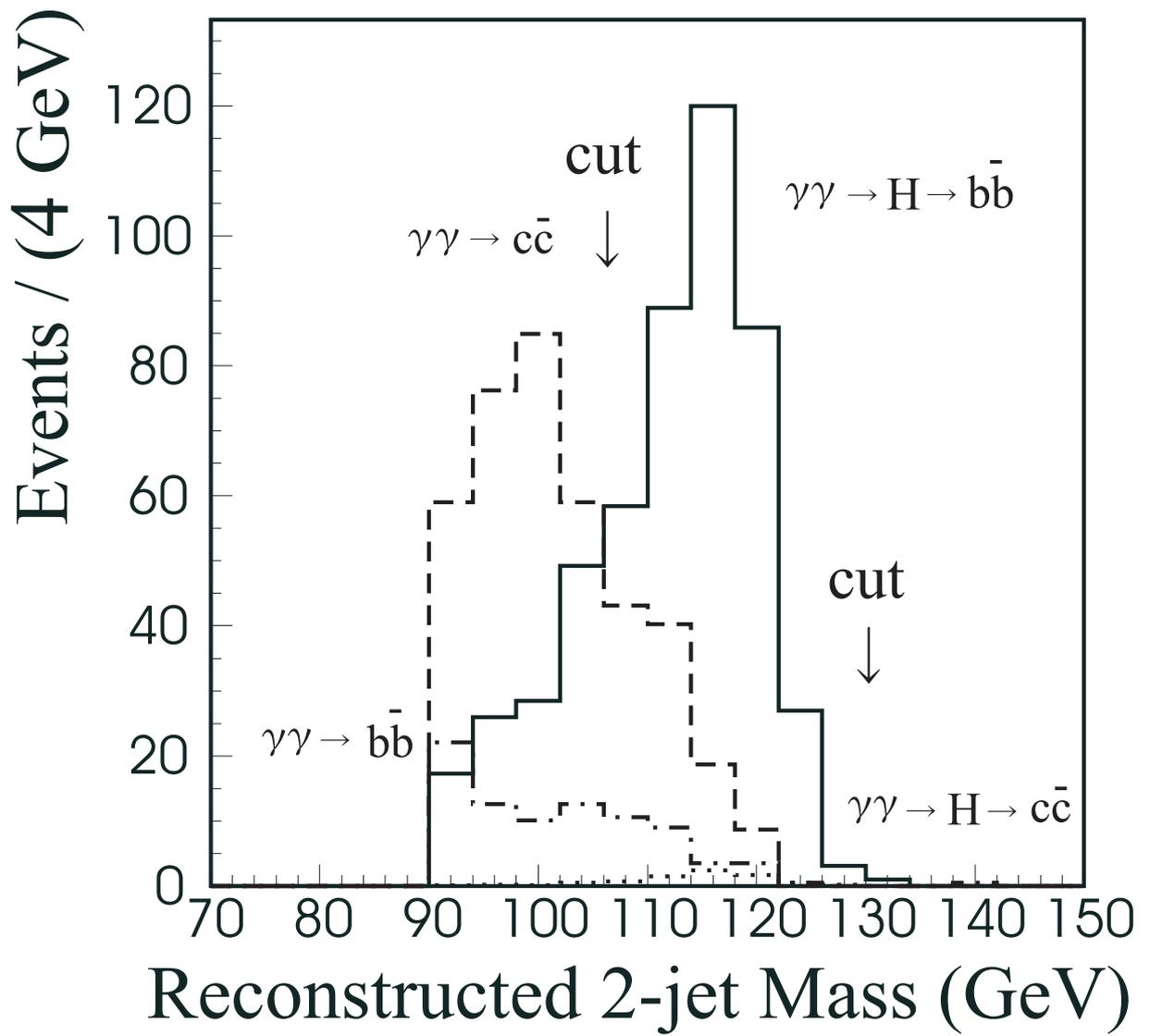
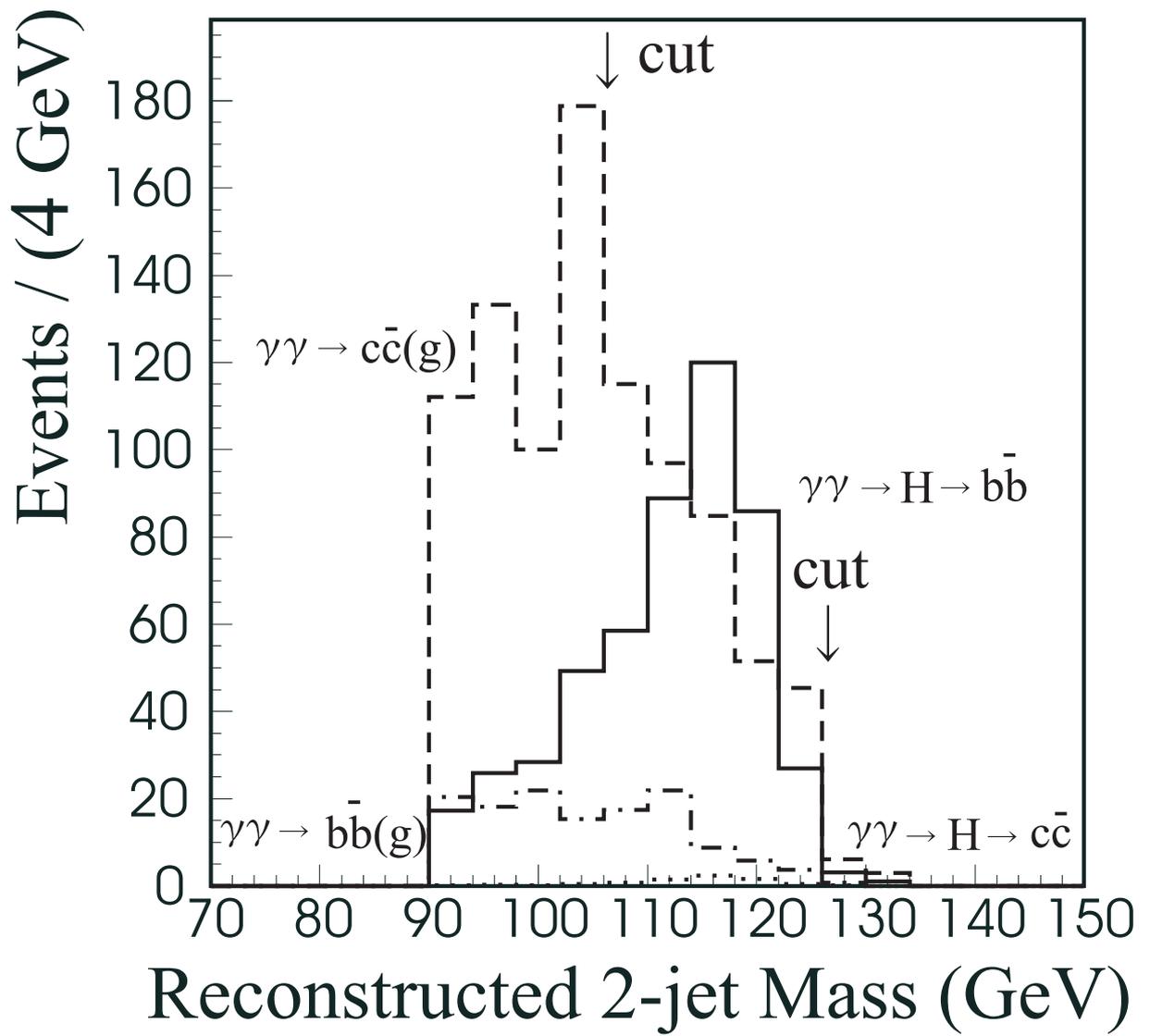
\begin{figure}
\caption{The reconstructed invariant mass distributions of two-jet
events with applying the $b$-tagging requirements. An integrated
luminosity of 10 $\mbox{fb}^{-1}$ and standard model branching
fractions for the Higgs boson are assumed. (a) The background events
are evaluated in the tree-level. (b) The effect of QCD
corrections to background cross sections is taken into account.}
\label{fig:higgs1}
\end{figure}

  To show the effect of QCD radiative corrections to the background 
processes explicitly, the distribution of the selected events against 
the reconstructed two-jet invariant mass $M_{jj}$ at the tree-level 
and with the QCD corrections are displayed in Fig.~\ref{fig:higgs1}, 
separately. In order to enhance the signal, a cut of the invariant
mass is tuned in a way that the statistical significance of the signal
over backgrounds, $(N_{obs}-\langle N_{bg}\rangle)/\sqrt{N}_{obs}$, is 
maximized, where $N_{obs}$ is the number of observed events and
$\langle N_{bg}\rangle$ is the number of expected background events. 
As a results, events in the two-jet mass ranges 106 GeV $< M_{jj} <$ 
130 GeV and 106 GeV $< M_{jj} <$ 126 GeV are adopted for the tree-level 
and QCD corrected evaluations of the backgrounds,
respectively. Table~\ref{tbl:sel1} and~\ref{tbl:sel2} list the number of
events in the invariant mass range, which are the final candidate
events as the $\gamma\gamma \to H \to b\bar{b}$. The numbers of
estimated signal and background are 383 and 146 from the tree-level 
computation, while 380 and 459 with the QCD corrections, respectively. 
Most of the backgrounds are from $\gamma\gamma\to c\bar{c}(g)$ process.
The selection efficiencies in the above invariant mass ranges, 
\begin{eqnarray}
\label{eq:6}
  \varepsilon_{sel}=\frac{\# \ \mbox{of selected events}}{\# \
\mbox{of generated events}}.
\end{eqnarray}
are listed in Table~\ref{tbl:sel1} and~\ref{tbl:sel2} for each cases.
The $b\bar{b}$-tagging efficiency in the accepted invariant mass 
range defined as,
\begin{eqnarray}
\label{eq:5}
  \varepsilon_{tag}=\frac{\# \ \mbox{of selected events}}{\# \ 
\mbox{of two-jet events in the mass range}}.
\end{eqnarray}
is also found in Table~\ref{tbl:sel1} and~\ref{tbl:sel2}.

  The two-photon decay width of the Higgs boson is proportional to 
the event rates of the Higgs signal. The statistical error of the
number of signal events $\sqrt{N}_{obs}/(N_{obs}-\langle
N_{bg}\rangle)$ directly corresponds to the statistical error of the 
measurement of the two-photon decay width, while the other origins of
the errors such as the background subtraction, luminosity
distribution, etc., influence the systematic
error. Table~\ref{tbl:res} lists the statistical errors of two-photon
decay width of the Higgs boson. The two-photon decay width of the SM
Higgs boson at $M_{H}$=120 GeV, in the estimate with the QCD
corrections to $\gamma\gamma \to q\bar{q}$ background processes, is
7.6\%. 
     

\section{Summary}

  We have studied feasibility of the measurement of two-photon decay
width of intermediate-mass Higgs boson in the standard model at a 
photon-photon collider by Monte Carlo simulations of photon-photon 
collisions, hadronizations and detector simulation. The QCD radiative 
corrections to the background process $\gamma\gamma \to q\bar{q}$ are 
taken into account. The statistical error on the measurement of the 
two-photon decay width of the Higgs boson with the mass of 120 GeV is 
7.6\% for the integrated luminosity of 10 $\mbox{fb}^{-1}$. At the
integrated luminosity of 20 $\mbox{fb}^{-1}$, the ratio of signal to 
background is improved to be 760/919, and the statistical errors on
the two-photon decay width measurement for 120 GeV Higgs boson is
5.4\%. 

  The statistical errors of the two-photon decay width of the
intermediate-mass Higgs boson using Monte Carlo simulation by Borden
et al.\cite{bor93} are within 5\% when the background events at
tree-level and the integrated luminosity 20 $\mbox{fb}^{-1}$ are
assumed. The statistical errors in our analysis are comparable
with their study. In \cite{bor93}, the $b\bar{b}$-tagging efficiency
is assumed to be 50\% with 5\% $c\bar{c}$ contamination, while it is
estimated to be 64.4\% with 15.1\% contamination by the detector
simulation in the present study. Since the adopted $b$-quark tagging
algorithm in our analysis is simple one in which the three dimensional
impact parameters are computed from the tracking data in the vertex
detector, it is expected that the developments of new tagging
algorithms and particle identification can be more efficient in
separating $H \to b\bar{b}$ events from other backgrounds.

  This result shows, for instance, that the photon-photon collider
will be sufficient to distinguish the intermediate-mass Higgs boson of
SM from the lightest Higgs of MSSM, if the ratio of the two-photon
decay widths $\Gamma(h^{0} \to \gamma\gamma,\mbox{MSSM})/
\Gamma(H \to \gamma\gamma, \mbox{SM})$ is as large as 1.2 \cite{bor93}.
It indicates that a photon-photon collider has a great and unique
feasibility to look for the new physics beyond SM.  


\section{Acknowledgments}

  We greatly appreciate Prof. I.~Endo for useful discussions and 
encouragement. We would like to thank Profs. G.~Jikia, J.~Kamoshita, 
T.~Kon, Y.~Okada, T.~Takeshita, T.~Tauchi, A.~Tkabladze, A.~Miyamoto 
and K.~Yokoya for useful discussions. We thank members of Photon
Physics Laboratory at Hiroshima University, and members of Akita
Junior College. This work was partly supported by the Grant-in-Aid for 
Scientific Research from Ministry of Education, Science and Culture of 
Japan. One of the authors, T.Ohgaki, would like to thank the Research 
Fellowships of the Japan Society for the Promotion of Science for
Young Scientists.


\newpage

\begin{figure}
\caption{The coupling of the Higgs boson with two photons generated by
a loop of massive charged particle.}
\label{fig:cou}
\end{figure}

\begin{figure}
\caption{Branching ratio of the standard model Higgs boson. The top
quark mass is assumed to be 176 GeV. Computed by HDECAY
\protect\cite{spi96}.}
\label{fig:bra}
\end{figure}

\begin{figure}
\caption{The polarized luminosity distributions of a photon-photon
collider at $\protect\sqrt{s}_{e+e-}$=150 GeV with
$P_{L}P_{e}=-1.0$. The bin size is 0.02. (a) $J_{z}=0$. (b)
$J_{z}=\pm2$.}
\label{fig:lum1}
\end{figure}
 
\begin{figure}
\caption{The luminosity distribution of a photon-photon collider at
$\protect\sqrt{s}_{e+e-}$=150 GeV with $P_{L}P_{e}=-1.0$ in the
$z$-$\eta$ plane. The vertical axis is $d^{2}L_{\gamma\gamma}/dzd\eta$
in units of $\mbox{nb}^{-1}\mbox{s}^{-1}/\mbox{\rm bin}$. The bin size
is 0.02$\times$0.08.}
\label{fig:lum2}
\end{figure}

\newpage

\mediumtext
\begin{table}
\caption{Parameters of the photon-photon collider based on JLC 
for $M_H$=120 GeV.}
\label{tbl:par}
\begin{tabular}{cccc}
{Electron beam parameters} & & & \\
\hline
  Number of electrons per bunch & $N_e$     &
$0.63\times10^{10}$ & \\
  Number of bunches per pulse   & $m_b$     & 85 & \\
  Repetition rate               & $f_{rep}$ & 150 & Hz \\
  Normalized emittance 
              & $\gamma\epsilon_{x,e}$ & $3.3\times10^{-6}$ & m
\\
              & $\gamma\epsilon_{y,e}$ & $4.8\times10^{-8}$ & m
\\
  R.m.s.~bunch length & $\sigma_{z,e}$ & 90 & $\mu$m \\
  Beta functions at I.P. & $\beta^{*}_{x,e}$ & 0.30 & mm \\
                         & $\beta^{*}_{y,e}$ & 10.0 & mm \\
  Beam size at I.P. without conversion 
             & $\sigma^{*}_{x,e}$ & 82 & nm \\
             & $\sigma^{*}_{y,e}$ & 57 & nm \\
  Beta functions at C.P. 
             & $\beta^{CP}_{x,e}$ & 0.33 & m \\
             & $\beta^{CP}_{y,e}$ & 20 & mm \\
  Beam size at C.P. 
             & $\sigma^{CP}_{x,e}$ & 2.7 & $\mu$m \\
             & $\sigma^{CP}_{y,e}$ & 81 & nm \\
\hline
{Laser beam parameters} & & & \\
\hline
  Wavelength        & $\lambda_L$ & 0.297 & $\mu$m \\ 
  Photon energy & $\hbar\omega_L$ & 4.18 & eV \\
  R.m.s.~pulse length & $\sigma_{z,L}$ & 300 & $\mu$m (1ps) \\
  Laser beam size at C.P.
             & $\sigma^{CP}_{x,L}$ & 5 & $\mu$m \\
             & $\sigma^{CP}_{y,L}$ & 5 & $\mu$m \\
  Number of laser photons in a pulse & $N_L$ &
$1.1\times10^{19}$ & \\
  Energy per pulse & $\hbar\omega_L N_L$ & 7 & Joule \\
  Laser peak power (effective rectangular pulse) & $P_L$ & 2.0 
& TW \\
  Maximum electric field (Gaussian peak) & ${\cal E}_{L,max}$
& $2.2\times10^{12}$ & V/m \\
  Nonlinear QED parameter at Gaussian peak & $\xi_{peak}$ & 0.20 &\\
\hline
{Photon beam} & & & \\
\hline
Number of photons per electron bunch & $N_{\gamma}$ &
$0.41\times10^{10}$ &\\
Beam size at I.P. & $\sigma^{*}_{x,\gamma}$ & 107 & nm \\
                  & $\sigma^{*}_{y,\gamma}$ & 89 & nm \\
$\gamma\gamma$ luminosity & ${\cal L}_{\gamma\gamma}$ & 
                  $3.4\times10^{32}$ & cm$^{-2}$s$^{-1}$ \\
Distance between C.P. to I.P. & $L$ & 1.0 & cm\\
\end{tabular}
\end{table}

\begin{table}
\caption{Effective cross sections and generated events at a
photon-photon collider. The continuum backgrounds are generated as
$\protect\sqrt{s}_{\gamma\gamma}>75$ GeV.}
\label{tbl:crs}
\begin{tabular}{lccc}
       & $\sigma^{eff}_{\vert\cos\theta\vert<0.95}$ & 
Number of events & Number of \\
       &  (pb) & (10 $\mbox{fb}^{-1}$) & simulated events \\
\hline
Signal events                               & & & \\
  \hspace*{3mm} $\gamma\gamma \to H \to b\bar{b}$  &  0.508 &   5080 & 10000 \\
Backgrounds                                 & & & \\
  \hspace*{3mm} $\gamma\gamma \to H \to c\bar{c}$  & 0.0210 &    210 & 10000 \\
  \hspace*{3mm} $\gamma\gamma \to H \to gg      $  & 0.0633 &    633 & 10000 \\
  \hspace*{3mm} $\gamma\gamma \to b\bar{b}$        &  0.502 &   5020 & 10000 \\
  \hspace*{3mm} $\gamma\gamma \to c\bar{c}$        &   7.19 &  71900 & 50000 \\
  \hspace*{3mm} $\gamma\gamma \to b\bar{b}(g)$     &  0.727 &   7270 & 10000 \\
  \hspace*{3mm} $\gamma\gamma \to c\bar{c}(g)$     &   15.1 & 151000 & 50000 \\
\end{tabular}
\end{table}

\begin{table}
\caption{Performance parameters of the JLC-I detector. The units of
energies and momenta are in GeV.}
\label{tbl:jlc}
\begin{tabular}{ll}
  Vertex detector (VTX)       & \\
  \hspace*{3mm} Position resolution      & $\sigma$= 7.2 $\mu$m \\
  \hspace*{3mm} Impact parameter resolution & $\sigma_{d}^{2}= 
11.4^{2}+(28.8/p)^{2}/\sin^{3}{\theta}$ ($\mu\mbox{m}^{2}$) \\ 
  Central drift chamber (CDC) & \\
  \hspace*{3mm} Position resolution         & $\sigma_{x}$= 
100 $\mu$m, $\sigma_{y}$= 2 mm \\
  \hspace*{3mm} Momentum resolution         & $\sigma_{Pt}/Pt$= 
1.1$\times10^{-4} Pt$ $\oplus$ 0.1\% \\
                              & $\sigma_{Pt}/Pt$= 5$\times10^{-5} 
Pt$ $\oplus$ 0.1\% \\
                              & (with vertex constraint) \\
  Electromagnetic calorimeter (EM) & \\
  \hspace*{3mm} Energy resolution           & $\sigma_{E}/E$= 
15\% / $\sqrt{E}$ $\oplus$ 1\% \\
  Hadron calorimeter (HAD)    & \\
  \hspace*{3mm} Energy resolution           & $\sigma_{E}/E$= 
40\% / $\sqrt{E}$ $\oplus$ 2\% \\
  Magnetic field              & 2.0 T \\
\end{tabular}
\end{table}

\begin{table}
\caption{The number of $b\bar{b}$ tagged events with
10$\mbox{fb}^{-1}$.}
\label{tbl:tag}
\begin{tabular}{lc}
                    & Events \\
\hline
Signal events                 &  \\
  \hspace*{3mm} $H \to b\bar{b}$            & 582 \\
Backgrounds                   & \\
  \hspace*{3mm} $H \to c\bar{c}$            & 7.85 \\
  \hspace*{3mm} $H \to gg      $            & 1.58 \\  
  \hspace*{3mm} $\gamma\gamma \to b\bar{b}$    & 185 \\
  \hspace*{3mm} $\gamma\gamma \to c\bar{c}$    & 715 \\
  \hspace*{3mm} $\gamma\gamma \to b\bar{b}(g)$ & 278 \\
  \hspace*{3mm} $\gamma\gamma \to c\bar{c}(g)$ & 1320 \\
\end{tabular}
\end{table}

\begin{table}
\caption{The number of the accepted events as candidates of
$\gamma\gamma \to H \to b\bar{b}$, selection efficiencies and
$b$-tagging efficiencies in each processes. The tree-level backgrounds 
are assumed. The invariant mass range 106 GeV $< M_{jj} <$ 130 GeV is
adopted.}
\label{tbl:sel1}
\begin{tabular}{lccc}
   & Events & $\varepsilon_{sel}$ \ (\%) & $\varepsilon_{tag}$ \ (\%) \\       \hline
Signal events                                & & & \\
  \hspace*{3mm} $\gamma\gamma \to H \to b\bar{b}$   & 383 & 7.54 & 64.4 \\
Backgrounds                                  & & & \\
  \hspace*{3mm} $\gamma\gamma \to H \to c\bar{c}$   & 6.79 & 3.23 & 20.3 \\
  \hspace*{3mm} $\gamma\gamma \to H \to gg$         & 1.46 & 0.230 & 5.82\\
  \hspace*{3mm} $\gamma\gamma \to b\bar{b}$         & 27.1 & 0.540 &
77.1 \\
  \hspace*{3mm} $\gamma\gamma \to c\bar{c}$         & 111 & 0.154 &
15.1 \\
\hline
Signal to Background at tree-level & 383 / 146 & & \\
\end{tabular}
\end{table}

\begin{table}
\caption{The number of the accepted events as candidates of
$\gamma\gamma \to H \to b\bar{b}$, selection efficiencies and
$b$-tagging efficiencies in each processes. The backgrounds with QCD
corrections are assumed. The invariant mass range 106 GeV $< M_{jj} <$
126 GeV is adopted.}
\label{tbl:sel2}
\begin{tabular}{lccc}
   & Events & $\varepsilon_{sel}$ \ (\%) & $\varepsilon_{tag}$ \ (\%)
\\
\hline
Signal events                                & & & \\
  \hspace*{3mm} $\gamma\gamma \to H \to b\bar{b}$   & 380 & 7.48 &
64.3 \\
Backgrounds                                  & & & \\
  \hspace*{3mm} $\gamma\gamma \to H \to c\bar{c}$   & 6.65 & 3.16 &
20.1 \\
  \hspace*{3mm} $\gamma\gamma \to H \to gg$         & 1.46 & 0.230 &
5.88 \\
  \hspace*{3mm} $\gamma\gamma \to b\bar{b}(g)$      & 57.4 & 0.790 &
69.9 \\
  \hspace*{3mm} $\gamma\gamma \to c\bar{c}(g)$      & 394 & 0.260 &
16.2 \\
\hline
Signal to Background with QCD corrections & 380 / 459 & & \\
\end{tabular}
\end{table}

\begin{table}
\caption{Statistical errors on the measurement of two-photon decay
width of SM Higgs boson with the mass of 120 GeV. An integrated
luminosity of 10 $\mbox{fb}^{-1}$ is assumed.}
\label{tbl:res}
\begin{tabular}{ccc}
  $ \triangle X / X $ & tree-level & QCD correction \\
\hline
  $\Gamma(H \to \gamma\gamma)$ & 6.0 \% & 7.6 \% \\
\end{tabular}
\end{table}



\begin{references}

\bibitem{gin81} I.F.~Ginzburg, G.L.~Kotkin, V.G.~Serbo, and
V.I.~Telnov, {\it Pisma Zh.\ Eksp.\ Teor.\ Fiz.} {\bf 34}, 514 (1981); 
{\it JETP Lett.} {\bf 34}, 491 (1982).

\bibitem{gin83} I.F.~Ginzburg, G.L.~Kotkin, V.G.~Serbo, and
V.I.~Telnov, {\it Nucl.\ Instrum.\ and Methods} {\bf 205}, 47 (1983); 
{\bf 219}, 5 (1984).

\bibitem{tel90} V.I.~Telnov, {\it Nucl.\ Instrum.\ and Methods} 
{\bf A294}, 72 (1990).

\bibitem{tel95} V.I.~Telnov, in {\it Proceedings of the Workshop on
Gamma-Gamma Colliders}, Berkeley, CA, USA, 1994,
eds.\ S.~Chattopadhyay {\it et al.}, {\it Nucl.\ Instrum.\ and 
Methods} {\bf A355}, 3 (1995).

\bibitem{wat93} Y.~Yasui, I.~Watanabe, J.~Kodaira, and I.~Endo,
{\it Nucl.\ Instrum.\ and Methods} {\bf A335}, 385 (1993).

\bibitem{bor92} D.L.~Borden, D.A.~Bauer, and D.O.~Caldwell,
SLAC preprint, SLAC-PUB-5715 (1992).

\bibitem{gun93} J.F.~Gunion and H.E.~Haber, {\it Phys.\ Rev.} {\bf D48},
5109 (1993).

\bibitem{bor93} D.L.~Borden, D.A.~Bauer, and D.O.~Caldwell, 
{\it Phys.\ Rev.} {\bf D48}, 4018 (1993).

\bibitem{bor94} D.L.~Borden, V.A.~Khoze, W.J.~Stirling, and 
J.~Ohnemus, {\it Phys.\ Rev.} {\bf D50}, 4499 (1994).

\bibitem{bai94} M.~Baillargeon, G.~B\'elanger, and 
F.~Boudjema, ENSLAPP preprint, ENSLAPP-A-473/94 (1994).

\bibitem{bai95} M.~Baillargeon, G.~B\'elanger, and 
F.~Boudjema, {\it Phys.\ Rev.} {\bf D51}, 4712 (1995).

\bibitem{wat95} I.~Watanabe, in {\it Proceedings of INS Workshop on 
the Physics of $e^{+}e^{-}$, $e^{-}\gamma$ and $\gamma\gamma$ 
Collisions at Linear Accelerators}, Tokyo, Dec. 20-22, 1994,
eds.\ Z.~Hioki, T.~Ishii, and R.~Najima, INS-J-181, 139 (1995).  

\bibitem{wat96} I.~Watanabe, in {\it Proceedings of Workshop on 
Physics and Experiments with Linear Colliders}, Morioka-Appi, Iwate,
Japan, Sep. 8-12, 1995, eds.\ A.~Miyamoto, Y.~Fujii, T.~Matsui, and
S.~Iwata, World Scientific, 689 (1996).

\bibitem{jik95} G.~Jikia and A.~Tkabladze, {\it Nucl.\ Instrum.\ and
Methods} {\bf A355}, 81 (1995).

\bibitem{jik96} G.~Jikia and A.~Tkabladze, {\it Phys.\ Rev.} {\bf
D54}, 2030 (1996).

\bibitem{nlc96} S.~Kuhlman et al., NLC ZDR Design Group and the NLC
Physics Working Group, {\it Physics and Technology of the Next Linear
Collider}, BNL 52-502, report submitted to Snowmass '96 (1996).

\bibitem{jik93} G.V.~Jikia, {\it Nucl.\ Phys.} {\bf B405}, 24 (1993).

\bibitem{naj89} R.~Najima, in {\it Proceedings of Third Meeting on
Physics at TeV energy Scale}, KEK, Sep. 28-30, 1989, eds.\ K.~Hidaka
and C.S.~Lim, KEK Report 90-9, 112 (1990). 

\bibitem{jon79} D.R.T.~Jones and S.T.~Petcov, {\it Phys.\ Lett.} {\bf
B84}, 440 (1979).

\bibitem{gun88} J.F.~Gunion, G.L.~Kane, and J.~Wudka, {\it Nucl.\
Phys.} {\bf B299}, 231 (1988).

\bibitem{gun91} J.F.~Gunion, {\it Phys.\ Lett.} {\bf B261}, 510
(1991).

\bibitem{kun91} Z.~Kunszt, Z.~Trocsanyi, and W.J.~Stirling, {\it
Phys.\ Lett.} {\bf B271}, 247 (1991).

\bibitem{gun90} J.F.~Gunion, H.E.~Haber, G.L.~Kane, and 
S.~Dawson, {\it The Higgs Hunter's Guide}, (Addison-Wesley, 
Redwood City, CA, 1990).

\bibitem{oku82} L.B.~Okun, {\it Leptons and Quarks}, (North-Holland, 
Amsterdam, 1982).

\bibitem{che95} P.~Chen, G.~Horton-Smith, T.~Ohgaki, 
A.W.~Weidemann, and K.~Yokoya, 
{\it Nucl.\ Instrum.\ Methods} {\bf A355}, 107 (1995).

\bibitem{ohg95} T.~Ohgaki and T.~Takahashi, {\it Nucl.\ Instrum.\
Methods} {\bf A373}, 185 (1996).

\bibitem{che96} P.~Chen, T.~Ohgaki, A.~Spitkovsky, T.~Takahashi, and
K.~Yokoya, in preparation. 

\bibitem{sjo94} T.~Sj\"{o}strand, {\it Comput. Phys. Commun.} {\bf
82}, 74 (1994).

\bibitem{jlc92} {\it JLC-I}, KEK Report 92-16, December (1992).

\bibitem{spi96} M.~Spira, in {\it Proceedings of 5th International Workshop 
on New Computing Techniques in Physics Research: Software Engineering, 
Neural Nets, Genetic Algorithms, Expert Systems, Symbolic Algebra,
Automatic Calculations (AIHENP 96)}, Lausanne, Switzerland, Sep. 2-6,
1996.

\bibitem{kaw86} S.~Kawabata, {\it Comput.\ Phys.\ Commun.} {\bf 41}, 127 
(1986); {\it ibid.} {\bf 88}, 309 (1995). 

\bibitem{mur92} H.~Murayama, I.~Watanabe, and K.~Hagiwara, 
KEK Report 91-11, January (1992).

\bibitem{bar84} JADE Collaboration, W.~Bartel {\it et al.}, {\it Z.\ Phys.} 
{\bf C26}, 93 (1984).

\end{references}
\end{document}